\documentclass[12pt]{article}

\usepackage{amsmath,amssymb,graphicx,latexsym}
\usepackage{hyperref}
\usepackage{cite}
\newcommand{\ed}{\end{document}}
\newcommand{\beq}{\begin{equation}}
\newcommand{\eeq}{\end{equation}}
\newcommand{\beqa}{\begin{eqnarray}}
\newcommand{\eeqa}{\end{eqnarray}}
\newcommand{\bc}{\begin{center}}
\newcommand{\ec}{\end{center}}

\newcommand{\ba}{\begin{array}}
\newcommand{\ea}{\end{array}}
\newcommand{\pa}{\partial}

\begin{document}
\title{\bf{Simulating gravity in rotational flow}}
\author
{Satadal Datta$^{1,2}$, Arpan Krishna Mitra$^1$\\
{\small\textit{ $^ {1}$ Harish-Chandra Research Institute, HBNI,
Chhatnag Road, Jhunsi, Allahabad 211 019, India}}\\
{\small\textit{$^2$ Department of Physics and Astronomy, Seoul National University, 08826 Seoul, Korea}}
\date{}
}
\maketitle

\begin{abstract}
We consider classical fluids in non-relativistic framework. The flow is considered to be barotropic, inviscid and rotational. We study the linear perturbations  over a steady state background flow.
 We find the acoustic metric from the conservation equation of a current constructed from linear perturbation of first order derivatives (in position and time coordinate) of Bernoulli's constant (scalar field) and vorticity (a vector field). We have rather shown that the conservation equation of current reduces to a massless scalar field equation in the high frequency limit. In contrast to the contemporary works, our work shows that even if we can not find a wave equation (in rotational flow) which is structurally similar to a massless scalar field equation in curved space-time, but still an analogue space-time exists through a conservation equation.   
 Considering velocity potential and Clebesch coefficients, we find that only for some specific systems current conservation equation can be found yielding the same analogue space-time. We conclude that for rotational flows, it is wise to study linear perturbation of Bernoulli's constant over the velocity potential and Clebsch coefficients.  
\end{abstract}
\pagebreak
\section{Introduction}
{\it Unruh's pioneering work,} Unruh's work \cite{unruh} demonstrated that in an irrotational (potential flow), barotropic, inviscid flow, linearisation (linearising fluid density and velocity) of the fluid equations over a background flow (a known solution of density and velocity of the fluid equations) result in a wave equation of the velocity potential. This wave equation is similar to that of a minimally coupled massless scalar field equation in curved space-time yielding pseudo-Riemanninan geometry. From this, one can construct a space-time geometry, namely the acoustic metric. Therefore, sound (linear perturbation in density and velocity) propagation is absolutely influenced by the background flow of the medium just like a massless scalar field's propagation depends on the background space-time geometry in General Relativity. Thus by cautiously designing the background flow of the medium, one can even simulate black-hole horizon space-time and calculate corresponding Hawking radiation and Hawking Temperature owing to the linear dispersion relation of the phonons (quanta of sound) in the medium. There have been many works in this subject \cite{barcelo}\cite{Visser_1998} which covers adiabatic equation of state too whereas Unruh's work considered isothermal flow. In this connection, there have been works where the irrorationality condition is relaxed to study analogue gravity in rotational fluids to simulate Kerr like space-time geometry \cite{Visser_2005} and also to study sound wave propagation in a rotational flow \cite{bergliaffa2004wave}. In all these works, the authors have done approximations in rotational flow while studying analogue gravity. To be precise in the work \cite{Visser_2005} the authors have worked in geometrical acoustic limit to find Kerr like space-time, and in the work \cite{bergliaffa2004wave}, the authors have used Pierce's approximation \cite{pierce} in the context of analogue gravity.\\
\indent 
Here we introduce linear perturbations in fluid density and velocity field over a steady background flow (not necessarily locally irrotational). We have found a current conservation equation from which the acoustic metric components can be derived. In this process, we have used no such aforementioned assumptions/approximations on the flow. Our only assumption requires the background flow to be steady. This is why, the resulting acoustic metric is found to be stationary. We have shown how physical acoustics and geometrical acoustics (the approximation useed in the paper \cite{Visser_2005}) are connected in the high frequency limit. We have shown the existence of an acoustic metric even when the Pierce approximation or geometrical acoustic limit (the assumptions considered so far in the literature) are not considered. We construct acoustic metric by considering a different quantity other than the velocity potential part in the rotational flow velocity. 
Further, we have modified the the definition of velocity to accommodate it with the vorticity present in the fluid. We have used the Clebsch parameters for this. The extra piece of terms present the linearised perturbation equation of the velocity potential can be interpreted as current terms for some real system in absence of radial force when the vorticity is small. 

In this paper we have introduced our work and discussed about the motivation behind it in the 1st section. Section 2 deals with the formulation using the Bernoulli's constant. In section 3 we have shown clearly that at the high frequency limit, the sound moving through the fluid can't see the vorticity present, in an elegant argumentative way. Section 4 is dedicated to the velocity potential formulation. We have concluded our discussion in section 5.
\section{Formulation}
In the work \cite{datta2018acoustic2}, the authors have considered Bernoulli's constant instead of velocity potential to study analogue gravity in the Astrophysical context. The potential flow was assumed to be inviscid and barotropic. The wave equation satisfied by the linear perturbation in Bernoulli's constant, $\zeta_1$ is as follows
\begin{equation}\label{bermoulli}
\partial_\mu (f^{\mu\nu}\partial_\nu\zeta_1)=0.
\end{equation}
The acoustic metric can be derived from $f^{\mu\nu}$ by comparing the wave equation with a minimally coupled scalar field: $f^{\mu\nu}=\sqrt{-g}g^{\mu\nu}$ in $3+1$D in general. We rewrite the Eq. \eqref{bermoulli} as 
\begin{equation}\label{current}
\partial_\mu \sqrt{-g}g^{\mu\nu}\partial_\nu\zeta_1=\partial_\mu \sqrt{-g} j^\mu=0,
\end{equation}
where $j^\mu=g^{\mu\nu}\partial_\nu\zeta_1=\partial^\mu\zeta_1$ (convention: raising upper index by $g^{\mu\nu}$) yields a current conservation equation in the analogue fluid model of gravity. Any massless scalar filed ($=\phi$) equation can be written as a current conservation equation in general, and it arises from the translational symmetry in $\phi$ in the action. The assumption of potential flow requires certain restrictions, symmetries in the flow. If the flow is not barotropic , it does not satisfies rotationality \cite{lagrangian}\cite{PhysRevD.96.064019}. For an oscillating body, if the amplitude of oscillation is less than the dimension of the oscillating body, flow past the body is irrotational in nature \cite{landau}. A body immersed in a irrotational (or potential) flow, flow near the body's surface is not vorticity free \cite{landau}.
\\
\indent
Here we make an improvement over the restrictions on rotationality, to achieve more generality. We consider rotational inviscid barotropic non-relativistic flow.
The fluid equations are given by 
\begin{eqnarray}
& \partial_t\rho+\nabla.(\rho{\bar v})=0, \label{continuity}\\
&\partial_t{\bar v}+{\bar v}.\nabla {\bar v}=-\frac{\nabla p}{\rho}-\nabla\psi,\label{euler}\\
& p=f(\rho),\label{barotropic}\\
& \nabla\times{\bar v}={\bar \omega}\label{rotationality}.
\end{eqnarray}
$\rho,~{\bar v}$ are fluid density and velocity respectively. $\psi$ is the potential corresponding to the external conservative force, it might be gravity or any trapping potential in general.
The above equations are continuity equation, Euler equation, barotropic condition and rotationality respectively. From Eq. \eqref{barotropic}, sound speed $c_s$ is given by
\begin{equation}
c_s^2=\frac{\partial p}{\partial \rho}=\frac{df}{d\rho}.
\end{equation}
 Taking a curl over the Eq. \eqref{euler} gives the famous equation of conservation of circulation \cite{landau}\cite{astrof}.
\\
From Eq. \eqref{euler}, we write
\begin{equation}\label{vJ}
\partial_t{\bar v}=-\tilde{{\bar J}},
\end{equation}
where 
\begin{eqnarray}
& \tilde{{\bar J}}=\nabla\zeta+{\bar \omega}\times{\bar v}, \label{J}\\
& \zeta=\frac{1}{2}{\bar v}^2+\int\frac{dp}{\rho}+\psi \label{z}.
\end{eqnarray}
We have the known background steady state solution $(\rho_0,{\bar v}_0)$ satisfying steady state fluid equations:
\begin{eqnarray}
&\nabla.(\rho_0{\bar v}_0)=0,\\
& {\bar v}_0.\nabla {\bar v}_0=-\frac{\nabla p}{\rho}|_{\rho=\rho_0}-\nabla\psi,\label{eulers}\\
&  p_0=f(\rho_0),\\
& \nabla\times{\bar v}_0={\bar \omega}_0.
\end{eqnarray}
Eq. \eqref{eulers} implies $\nabla\zeta_0=-{\bar \omega}_0\times{\bar v}_0\Rightarrow {\bar J}_0=0$, also evident from the Eq. \eqref{vJ} (${\bar J}_0$ is $\tilde{{\bar J}}$ for the background solution). 
Thermodynamic sound speed, $c_{s0}$, $c_{s0}^2=\frac{df}{d\rho}|_{\rho=\rho_0}$.
We introduce linear perturbations over this known solution, given by 
\begin{eqnarray}
& \rho=\rho_0+\epsilon\rho_1,\\
& {\bar v}={\bar v}_0+\epsilon {\bar v}_1;\\
&\Rightarrow p=p_0+\epsilon p_1,\\
& {\bar \omega}={\bar \omega}_0+\epsilon{\bar \omega}_1,\\
&\tilde{{\bar J}}={\bar J}_0+\epsilon{\bar J}_1,\\
& \Rightarrow \zeta_1={\bar v}_0.{\bar v}_1+\frac{c_{s0}^2\rho_1}{\rho_0}.
\end{eqnarray}
$\epsilon$ is a small quantity denoting small amplitude sound wave (see about more details in \cite{barcelo}). Therefore,
\begin{equation}
\partial_t{\bar v}_1=-{\bar J}=-\left(\nabla\zeta_1+{\bar \omega}_1\times{\bar v}_0+{\bar \omega}_0\times{\bar v}_1\right). \label{j1}
\end{equation}
Taking a partial time derivative in $\zeta_1$,
\begin{equation}
\frac{\rho_0}{c_{s0}^2}\partial_t\zeta_1=\frac{\rho_0}{c_{s0}^2}{\bar v}_0.\partial_t{\bar v}_1+\partial_t\rho_1. \label{z1}
\end{equation}
Linearised continuity equation is given by
\begin{equation}
\partial_t\rho_1+\nabla.(\rho_0{\bar v}_1+\rho_1{\bar v}_0)=0. \label{c1}
\end{equation}
Now taking a partial time derivative in Eq. \eqref{z1}, and using Eq. \eqref{j1}- Eq. \eqref{c1}, we have
\begin{equation}\label{fmn}
\partial_\mu f^{\mu\nu}J_\nu=0,
\end{equation}
where 
\begin{equation}
J_\mu\equiv\{\partial_t\zeta_1,{\bar J}\},
\end{equation}
and 
\begin{equation}
f^{\mu\nu}\equiv\frac{\rho_{0}}{c_{s0}^{2}}\begin{bmatrix}
-1 & \vdots & -{\bar v}_{0} \\
\cdots&\cdots&\cdots\cdots \\
-{\bar v}_{0}^T&\vdots & c_{s0}^{2}\mathbb{I}-{\bar v}_{0}\otimes{\bar v}_{0}
\end{bmatrix}
\end{equation}
Comparing Eq. \ref{fmn} to that of a current conservation equation in curved space-time, we set $f^{\mu\nu}=\sqrt{-g}g^{\mu\nu}$ yielding the determinant of the non singular matrix $g_{\mu\nu}$ (inverse of $g^{\mu\nu}$), $g=-\frac{\rho_0^4}{c_{s0}^2}$ in $3+1$ D. Hence we have
\begin{equation}\label{conservation1}
J^\mu~;\mu=\frac{1}{\sqrt{-g}}\partial_{\mu}\sqrt{-g}J^{\mu}=0,
\end{equation}
$J^\mu=g^{\mu\nu}J_{\nu}$, is given by
\begin{equation}
J^{\mu}=g^{\mu\nu}J_\nu=\left\{J^t\left[=\partial^t\zeta_1+\frac{1}{\rho_0 c_{s0}}{\bar v}_0.\left({\bar v}_1\times{\bar \omega}_0+{\bar v}_0\times{\bar \omega}_1\right)\right], {\bar J}\left[=J^t{\bar v}_0-\frac{c_{s0}}{\rho_0}\left({\bar v}_1\times{\bar \omega}_0+{\bar v}_0\times{\bar \omega}_1\right)\right]\right\},
\end{equation} 
the acoustic metric components are given by
\begin{equation}
g^{\mu\nu}\equiv\frac{1}{\rho_{0}c_{s0}}\begin{bmatrix}
-1 & \vdots & -{\bar v}_{0} \\
\cdots&\cdots&\cdots\cdots \\
-{\bar v}_{0}^{T}&\vdots & c_{s0}^{2}\mathbb{I}-{\bar v}_{0}\otimes{\bar v}_{0}
\end{bmatrix}
\end{equation}
\begin{equation}
\Rightarrow g_{\mu\nu}\equiv\frac{\rho_{0}}{c_{s0}}\begin{bmatrix}
 -(c_{s0}^{2}-v_{0}^{2}) & \vdots & -{\bar v}_{0} \\
\cdots&\cdots&\cdots\cdots \\
-{\bar v}_{0}^T&\vdots &\mathbb{I}
\end{bmatrix}
\end{equation}
The conservation equation is in harmony with the nondissipative nature of the system. 
Acoustic metric interval can be expressed as
\begin{equation} \label{ds2}
ds^{2}=\frac{\rho_{0}}{c_{s0}}\left[-(c_{s0}^{2}-v_{0}^{2})dt^{2}-{\bar{v}}_0.d{\bar{x}}dt-{\bar{v}}_0.dtd{\bar{x}}+d{\bar{x}}^{2} \right],
\end{equation}
where $v_0$ is the magnitude of the vector, ${\bar{v}}_0$. This metric has the same mathematical form with that for a potential flow \cite{barcelo}\cite{datta2018acoustic2}. The only difference is in the rotationality. This is how, we construct acoustic metric from a conservation equation instead of massless scalar field equation. It is not possible to find a massless scalar field without using Pierce approximation used in the work \cite{bergliaffa2004wave}. Here, the artificial gravity manifests itself through a conservation equation, Eq. \eqref{conservation1}. Eq. \eqref{conservation1} reduces to Eq. \eqref{current} for a irrotational flow,
\begin{equation}
J_\mu=j_\mu+\Xi_\mu,
\end{equation}
where 
\begin{equation}
\Xi_{\mu}=\left\{\Xi_t(=0),{\bar\Xi}\left(={\bar \omega}_1\times{\bar v}_0+{\bar \omega}_0\times{\bar v}_1\right)\right\}.
\end{equation}
Evidently $\Xi_\mu$ is zero for an irrotational flow.
\section{Role of Rotationality}
Using Eq. \eqref{euler} and Eq. \eqref{rotationality}, we have
\begin{equation}\label{omega}
\partial_t{\bar v}+{\bar \omega}\times{\bar v}=-\nabla\zeta
\end{equation}
In a rotating frame with angular velocity ${\bar \omega'}$, this equation takes form \cite{landau}
\begin{equation}\label{nI}
\partial_t{\bar v}+{\bar \omega}\times{\bar v}+2{\bar \omega'}\times{\bar v}=-\nabla\zeta-\frac{1}{2}\nabla({\bar \omega^{'}}\times{\bar r})^2
\end{equation}
If we imagine this rotating frame at each point (the term due to centrifugal force vanishes) in the fluid with angular velocity ${\bar\omega}'=-\frac{\omega}{2}$, the Eq. \eqref{nI} takes form similar to a potential flow locally. Thus, the term in the Euler equation due to vorticity plays the role of the Coriolis force in a rotating frame, i.e., in the flow as if the velocity field has a rotation at each point, with rotational angular velocity being $\frac{{\bar \omega}}{2}$.\\
\indent 
For a linear perturbation with frequency $\Omega$, $|\partial_t {\bar v}|\sim\Omega |{\bar v}|$, $|{\bar \omega}\times {\bar v}|\sim\omega v$, where $v=|{\bar v}|,\omega=|{\bar \omega}|$. Therefore, in the Eq. \eqref{omega}, comparing first two terms in the LHS, we see that if $\Omega>>\omega$, the role due to the vorticity becomes negligible. One can safely work in the irrotational limit. Vorticity is transparent to high frequency sound. Therefore, in this case, the additional terms due to vorticity in the current do not occur. In this high-frequency approximation, Eikonal approximation \cite{barcelo} is valid, and this yields geometrical acoustics regime, i.e., where the sound wave can be considered as a ray propagating along the null geometric, implying
\begin{equation}
ds^2=0.
\end{equation}
Therefore, any conformal factor in $g_{\mu\nu}$ do not matter. Thus, one can construct acoustic metric without conformal factor in the geometrical acoustic regime; given by
\begin{equation} \label{ds2g}
ds^{2}|_{\rm geometric}=\left[-(c_{s0}^{2}-v_{0}^{2})dt^{2}-{\bar{v}}_0.d{\bar{x}}dt-{\bar{v}}_0.dtd{\bar{x}}+d{\bar{x}}^{2} \right].
\end{equation}
Surprisingly, in the high frequency limit, the metric remains the same. This metric gives interesting space-time geometry discussed in details \cite{Visser_2005}. In general, a steady state flow gives rise to nonstatic stationary space-time geometry and in fact space-time metric similar to Kerr geometry can be constructed with rotational flow (local vorticity is nonzero) having zero radial velocity \cite{Visser_2005}.
\section{Clebsch parametrization}

While dealing with rotational fluids we have seen the linearly perturbed velocity potential satisfies the massless scalar field equation. We wonder how the things will be modified if we extend the definition of velocity so that it includes the vorticity effect. The simplest way is to express it in terms of three scalars namely, Clebsch parameters. The expression of the fluid velocity is now,
\begin{equation}
\bar{v}={{\nabla}}\phi + \beta{{\nabla}}\gamma.
\end{equation}
Where, vorticity $\bar\omega$ is, ${{\nabla}}\beta \times{{\nabla}}\gamma$.

With this definition of velocity we would like to write down the Euler equation, \eqref{euler}

\begin{equation}
\label{agrc1}
\pa_{t}({{\nabla}}\phi + \beta{{\nabla}}\gamma)+ \frac{1}{2}\nabla({{\nabla}}\phi + \beta{{\nabla}}\gamma)^{2}+\nabla h(\rho)+ \nabla \psi = \bar v\times\bar\omega,
\end{equation}
where, $\nabla h(\rho)=\frac{\nabla p}{\rho}$.

For a fluid flow which is purely anguler in nature and possesses cylindrical symmetry \cite{Visser_2005} we can show $$\nabla\times(\bar v\times\bar\omega)=0$$ when the system is void of any source or sink. For such a system we would like to replace the r.h.s of \eqref{agrc1} with the gradient of a scalar term, namely, $\alpha$.
\begin{equation}
\label{agrc2}
\bar v\times\bar \omega =\nabla \alpha.
\end{equation}
 Which produces,
\begin{equation}
\label{agrc3}
\pa_{t}({{\nabla}}\phi + \beta{{\nabla}}\gamma)+ \frac{1}{2}\nabla({{\nabla}}\phi + \beta{{\nabla}}\gamma)^{2}+\nabla h(\rho)+ \nabla \psi = \nabla\alpha
\end{equation}
We would confine our discussion to fluids with small vorticity which will be reflected in the calculation as we would neglect the terms which have more than two $\beta$ or $\gamma$ terms multiplied. Moreover, we are keeping the time evolution of the vorticity terms off the table for the time being.

With these considerations we get,
\begin{equation}
\label{agrc4}
\pa_{t}\phi +\frac{1}{2}(\nabla\phi)^{2}+ \nabla\phi .(\beta\nabla\gamma) +h(\rho)+\psi=\alpha
\end{equation}

Linearised Euler equation now looks like,
\begin{equation}
\label{agrc5}
\pa_{t}\phi_{1} +\bar v_{0}.(\nabla \phi_{1}+\beta_{1}\nabla\gamma_{0}+\beta_{0}\nabla\gamma_{1})+ \frac{c_{s0}^{2}\rho_{1}}{\rho_{0}}-\alpha_{1}=0.
\end{equation}
where, $h_{1}=\frac{1}{\rho_{0}}\frac{\pa p}{\pa \rho}|_{\rho_{0}}\rho_{1}$
Hence,
\begin{equation}
\label{agrc6}
\rho_{1}=-c_{s0}^{-2}\rho_{0}[\pa_{t}\phi_{1} +\bar v_{0}.(\nabla \phi_{1}+\beta_{1}\nabla\gamma_{0}+\beta_{0}\nabla\gamma_{1})-\alpha_{1}]
\end{equation}

by taking time derivative of \eqref{agrc6} and using linearised continuity equation \eqref{c1} , we arrive at the equation,

\begin{eqnarray}
\nonumber
\pa_{\mu}(f^{\mu\nu}\pa_{\nu}\phi_{1})= \pa_{t}(c_{s0}^{-2}\rho_{0}\alpha_{1})-\pa_{t}(\bar v_{0}c_{s0}^{-2}\rho_{0}(\beta_{1}\nabla\gamma_{0}+\beta_{0}\nabla\gamma_{1}))-\nabla[\bar v_{0}\bar v_{0} c_{s0}^{-2}\rho_{0}\\
(\beta_{1}\nabla\gamma_{0}+\beta_{0}\nabla\gamma_{1})]+\nabla.[\rho_{0}(\beta_{1}\nabla\gamma_{0}+\beta_{0}\nabla\gamma_{1})]+\nabla.[\bar v_{0} c_{s0}^{-2}\rho_{0}\alpha_{1}]
\label{agrc7}
\end{eqnarray}
Now, if we consider that only the zeroth order terms of the additional clebsch parameters are non zero, we end up with a   conservation equation,  

\begin{equation}
\pa_{\mu}{\sqrt{-g}}(g^{\mu\nu}\pa_{\nu}\phi_{1}+j'^{\mu})=0\rightarrow \frac{1}{{\sqrt{-g}}}\pa_{\mu}{\sqrt{-g}}({\it{J'}}^{\mu})=0
\end{equation}

 where $j'^{0}=-\frac{\rho_{o}}{\sqrt{-g} c_{s0}^{2}}\alpha_{1}$, $\bar{j}'=-\frac{\rho_{o}v_{0}}{\sqrt{-g}c_{s0}^{2} }\alpha_{1} $ and ${\it{J}'}^{\mu}=\pa^{\mu}\phi_{1}+j'^{\mu}$.
 
 The vorticity in fluid, even with a very small magnitude can produce some terms which can behave as current terms.
\section{Summary and Conclusions}
We have shown that an analogue space-time geometry exists for rotational flows even for rotational sound of low frequency. The dispersion is not linear at the low frequency\cite{bergliaffa2004wave}, therefore it is not possible to construct a wave equation similar to massless scalar field equation in Lorentzian geometry. We have demonstrated that an analogue space-time geometry emerges through a current conservation equation. The effect of vorticity in the flow is similar to the role of Coriolis force in a fluid in a rotational reference frame. In the high frequency limit, if the contribution from frequency is large compared to that due to vorticity, the current conservation equation reduces to a wave equation of Bernoulli's constant (similar to massless scalar field in minimal coupling).  The acoustic metric components remains the same in the high frequency limit. Our work shows that vorticity in a flow does not modify the analogue space-time, instead analogue gravity manifests itself in a different fashion, i.e., through the conservation of a current. In the article \cite{silke}, the authors have experimentally demonstrated that when wave propagates on the
surface of water it gets amplified after being scattered by a
draining vortex, thus an effect analogous to superradiance is observed. In this light, we have provided a theoretical extension by relaxing different restrictions in flow on the subject, analogue gravity in rotational flows. 

\bibliographystyle{ieeetr}
\bibliography{paponarpan}

\end{document}